\documentclass[preprint,12pt]{elsarticle}
\newtheorem{theorem}{Theorem}
\newtheorem{lemma}[theorem]{Lemma}
\newdefinition{rmk}{Remark}
\newproof{pf}{Proof}
\newproof{pot}{Proof of Theorem \ref{thm2}}
\newtheorem{corollary}[theorem]{Corollary}

\usepackage{amssymb}
\usepackage{amsmath}
\usepackage{graphicx}
\usepackage{mathtools}
\usepackage{algorithmicx}
\usepackage{algorithm}
\usepackage{algpseudocode}
\usepackage{float}
\usepackage{bm}
\numberwithin{figure}{section}

\journal{Nuclear Physics B}
\makeatletter
\def\ps@pprintTitle{%
 \let\@oddhead\@empty
 \let\@evenhead\@empty
 \let\@oddfoot\@empty
 \let\@evenfoot\@empty
}
\makeatother
\begin{document}

\begin{frontmatter}



\title{FPT Constant  Approximation Algorithms for Colorful Sum of Radii\footnote{This paper is dedicated to the memory of Professor Costas Iliopoulos. This research was supported by PIFI grant, Chinese Academy of Sciences President’s International Fellowship Initiative, Grant No.2025PG0005.}}


\author[a,b]{Shuilian Liu} 
\author[c]{Gregory Z. Gutin \corref{cor1}}
\cortext[cor1]{Corresponding author. Email: g.gutin@rhul.ac.uk}
\author[a,b]{Yicheng Xu}
\author[a,b]{Yong Zhang}
\affiliation[a]{organization={Shenzhen Institutes of Advanced Technology, Chinese Academy of Sciences},
            city={Shenzhen},
            postcode={518055}, 
            country={China}}
\affiliation[b]{organization={University of Chinese Academy of Sciences},
            city={Beijing},
            postcode={101408}, 
            country={China}}
\affiliation[c]{organization={Royal Holloway, University of London},
            city={Egham, Surrey},
            postcode={TW20 0EX}, 
            country={UK}}

\begin{abstract}
We study the colorful sum of radii problem, where the input is a point set $P$ partitioned into classes $P_1, P_2, \dots, P_\omega$, along with per-class outlier bounds $m_1, m_2, \dots, m_\omega$, summing to $m$. The goal is to select a subset $\mathcal{C} \subseteq P$ of $k$ centers and assign points to centers in $\mathcal{C}$, allowing up to $m_i$ unassigned points (outliers) from each class $P_i$, while minimizing the sum of cluster radii. The radius of a cluster is defined as the maximum distance from any point in the cluster to its center. The classical (non-colorful) version of the sum of radii problem is known to be NP-hard, even on weighted planar graphs. The colorful sum of radii is introduced by Chekuri et al. (2022), who provide an $O(\log \omega)$-approximation algorithm. In this paper, we present the first constant-factor approximation algorithms for the colorful sum of radii running in  FPT (fixed-parameter tractable) time. Our contributions are twofold: We design an iterative covering algorithm that achieves a $(2+\varepsilon)$-approximation with running time exponential in both $k$ and $m$; We further develop a $(7+\varepsilon)$-approximation algorithm by leveraging a colorful $k$-center subroutine, improving the running time by removing the exponential dependency on $m$.
\end{abstract}




\end{frontmatter}



\section{Introduction}
Centroid-based $k$-clustering problems have been extensively studied in theoretical computer science, combinatorial optimization, and computational geometry. Given a set $P$ of $n$ nodes in an edge-weighted complete graph and an integer $k$ (the weight of each edge is set to be the distance between its nodes), the task is to select a subset $\mathcal{C} \subseteq P$ of $k$ centers and assign each node to its nearest center, aiming to minimize a clustering cost function such as $k$-center, $k$-median, or $k$-means. The $k$-center problem seeks to minimize the maximum distance from any node to its assigned center (i.e., the maximum cluster radius), while the $k$-median problem minimizes the sum of distances from all nodes to their respective centers. A compromise objective is the sum of radii problem, which is also a sum-based objective, but minimizes the sum of the radii of the $k$ clusters. This objective is often favorable to $k$-center, as it helps to avoid the so-called dissection effect \cite{hansen1997cluster,monma1989partitioning}, a phenomenon in which spatially close points are assigned to different clusters. 

While previously less explored compared to classical clustering objectives, the sum of radii has become popular in recent years, especially in the design of approximation and FPT algorithms. The problem is known to be NP-hard even on weighted planar graphs and constant doubling dimension metrics, as shown by Gibson et al. \cite{gibson2010metric}. For general metrics, they also provide a quasi-polynomial time $(1+\varepsilon)$-approximation. Using a primal-dual technique, Buchem et al. \cite{buchem20243+} obtain a $(3+\varepsilon)$-approximation, improving upon earlier results with approximation ratios of $3.504$ \cite{charikar2004clustering} and $3.389$ \cite{friggstad2022improved}. On the parameterized side, Chen et al. \cite{chen2024parameterized} first present an FPT (fixed-parameter tractable) 2-approximation algorithm for the sum of radii problem, by a radius profile guessing scheme. Similar FPT techniques have also been applied to capacitated variants of the problem \cite{inamdar2020capacitated,bandyapadhyay2019constant,jaiswal2024fpt,filtser2024fpt}.

In recent years, fair clustering has emerged as a promising research direction aimed at ensuring that clustering outcomes do not disproportionately neglect or overrepresent certain groups. A foundational contribution in this area is due to Chierichetti et al.~\cite{chierichetti2017fair}, who initiated the study of fairness in clustering by proposing the notion of balanced clustering, which enforces proportional representation of groups within each cluster. Following this milestone, fair clustering has rapidly evolved into a diverse field with multiple fairness definitions, including bounded representation fairness \cite{bera2019fair}, proportionally fair clustering \cite{chen2019proportionally}, and socially fair clustering \cite{ghadiri2021socially}, and others. Research on fairness with respect to the sum of radii remains relatively recent. Carta et al.~\cite{carta2024fpt} present a $(6 + \varepsilon)$-approximation algorithm for the fair sum of radii problem under balanced clustering. More recently, Nezhad et al.~\cite{nezhad2025polynomial} 
develop a polynomial-time constant-factor approximation algorithm for this problem by leveraging a degree-constrained subgraph computation technique \cite{gabow1983efficient}.

We focus on a related but distinct model known as colorful clustering, introduced by Bandyapadhyay et al.~\cite{bandyapadhyay2019constant}. The input is a node set $P = \bigcup_{i=1}^\omega P_i$, an integer $k$, and a vector $\boldsymbol{m} = (m_1, \ldots, m_\omega)$ that bounds the number of outliers allowed per group. The goal is to select $k$ centers and assign each node to a center or leave it as an outlier, with at most $m_i$ outliers from each group $P_i$. When $\omega = 1$, the problem reduces to the classical outlier-robust clustering, originally introduced by Charikar et al. \cite{charikar2001algorithms}. In addition, the outlier-robust variant of the sum of radii problem admits a 3-approximation algorithm \cite{buchem20243+}, which remains the best-known result for this objective. Unlike traditional outlier-robust clustering, which may completely exclude certain groups as outliers, colorful clustering imposes multi-outlier constraints on each group to prevent this phenomenon. 

This formulation has recently attracted significant interest, leading to a range of approximation algorithms for colorful clustering under different objectives. Bandyapadhyay et al.~\cite{bandyapadhyay2019constant} show that the natural LP for colorful $k$-center has an unbounded integrality gap. They address this by constructing a simplified LP and exploiting properties of basic feasible solutions to design a 2-pseudo approximation algorithm. 
Jia et al. \cite{jia2022fair} design an ingenious rounding framework that builds on a 2-pseudo approximation algorithm. By carefully exploiting the structure of well-separated instances where optimal clusters are sufficiently distant, they convert the pseudo-solution into a true 3-approximation. Meanwhile, Anegg et al. \cite{anegg2022technique} integrate polyhedral sparsity arguments, based on Bandyapadhyay et al. \cite{bandyapadhyay2019constant}, with dynamic programming in a round-or-cut framework, resulting in a 4-approximation. Agrawal et al.~\cite{agrawal2023clustering} utilize a coreset for the $k$-median with outliers to develop an FPT $(1 + 2/e + \varepsilon)$-approximation algorithm for the colorful $k$-median. For the colorful sum of radii, Chekuri et al. \cite{chekuri2022algorithms} use the primal-dual method to design an \(O(\log \omega)\)-approximation algorithm. 
\paragraph{Our contributions} Recent advances in fixed-parameter tractable algorithms have shown their potential to overcome long-standing barriers in approximation algorithms. This work provides the first constant FPT approximation algorithms for the colorful sum of radii problem.  We propose an iterative covering algorithm that achieves a $(2+\varepsilon)$-approximation with running time exponential in both $k$ and the number of outliers $m$. It also directly applies to the outlier-robust sum of radii problem. Moreover, we develop a general framework that leverages any $\beta$-approximation algorithm for the colorful $k$-center problem, yielding a $(2\beta + 1 + \varepsilon)$-approximation for the colorful sum of radii. In particular, applying this framework to the best-known 3-approximation algorithm for colorful $k$-center yields a $(7+\varepsilon)$-approximation algorithm with running time exponential in $k$. 
\section{Preliminaries}
In the colorful clustering setting, we are given a set of points $P = \bigcup_{i=1}^\omega P_i$ in a metric space, where each $P_i$ is a class of size $ n_i = |P_i| $, and  $\boldsymbol{n} = (n_1, \ldots, n_\omega) \in \mathbb{N}^\omega$.  We are also given an integer $k$, a distance function $d$, and a multi-outlier vector $\boldsymbol{m} = (m_1, \ldots, m_\omega) \in \mathbb{N}^\omega$, where $m = \sum_{i=1}^\omega m_i$ is the total number of allowed outliers. Let $\boldsymbol{\rho} = \boldsymbol{n} - \boldsymbol{m}$ denote the minimum number of points to be covered from each class. The goal is to select a set of $k$ centers $\mathcal{C} \subseteq P$, an outlier set $\mathcal{M} \subseteq P$, and an assignment function $\sigma: P \to \mathcal{C} \cup \mathcal{M}$, such that at most $m_i$ points from $P_i$ are assigned to $\mathcal{M}$, i.e., $|\mathcal{M} \cap P_i| \leq m_i$ for all $i \in [\omega]$. The objective (colorful sum of radii) is to minimize
$$
\sum_{c \in \mathcal{C}} \max_{p \in \sigma^{-1}(c)} d(p, c).
$$

Let $\mathcal{O}^* = (\mathcal{C}^*, \mathcal{M}^*, \sigma^*)$ denote an optimal solution to an instance $\mathcal{I} = (P, d, k, \boldsymbol{m})$, where $\mathcal{C}^* = \{c_1^*, \ldots, c_k^*\}$. For each center $c_i^*$, define its cluster as
$
C_i^* = \{p \in P \setminus \mathcal{M}^* : \sigma^*(p) = c_i^*\},
$
and its radius $r_i^* = \max_{p \in C_i^*} d(p, c_i^*)$. Without loss of generality, assume the radii are ordered as $r_1^* \geq \cdots \geq r_k^*$, and let $\mathrm{OPT} = \sum_{i=1}^k r_i^*$ denote the optimal cost. A candidate solution can be described by a set of balls $\mathcal{B} = \{B(c_i, r_i)\}_{i=1}^k$, where $B(c, r) = \{p \in P : d(p, c) \leq r\}$, and its cost as $\mathrm{cost}(\mathcal{B}) = \sum_{i=1}^k r_i$.

It is well-known that the optimal radii for the sum of radii problem can be approximated within a $(1+\varepsilon)$ factor in FPT time with respect to $k$, and we extend this to the colorful sum of radii in the following lemma. A related approach appears in \cite{carta2024fpt}.
\begin{lemma}
\label{guess}
For any \(\varepsilon > 0\), there exists an algorithm that constructs a collection of near-optimal radius profiles in time 
$O\left(\log^k_{1+\varepsilon} \left(k/{\varepsilon}\right)\right),$ where \emph{near-optimal} means that we can compute a radius profile  \((\tilde{r}_1, \tilde{r}_2, \dots, \tilde{r}_k)\) satisfying 
\[
r_i^* \leq \tilde{r}_i \leq (1+\varepsilon) r_i^* \quad \text{for all } i \in [k],
\]
where \((r_1^*, r_2^*, \dots, r_k^*)\) represent the radius profile of an optimal solution to the colorful sum of radii.
 \end{lemma}
 \begin{pf}
Let $r$ denote the value of a $\beta$-approximate solution to the colorful $k$-center. Then the largest optimal radius $r_1^*$ lies within the interval $\left[\frac{r}{\beta}, k r \right].$
We partition this interval into smaller intervals defined as $\left[(1+\varepsilon)^{\ell-1} \frac{r}{\beta}, (1+\varepsilon)^\ell \frac{r}{\beta}\right]$, for $ \ell = 1, 2, \ldots, \left\lceil \log_{1+\varepsilon} (\beta k) \right\rceil.$ Since $r_1^*$ must fall into one of these intervals, suppose it lies in $\left[(1+\varepsilon)^{j-1} \frac{r}{\beta}, (1+\varepsilon)^j \frac{r}{\beta}\right]$. Selecting the right endpoint of this interval as our guess $\tilde{r}_1 = (1+\varepsilon)^j \frac{r}{\beta}$, we ensure that $r_1^* \leq \tilde{r}_1 \leq (1+\varepsilon) r_1^*$. Therefore, $\tilde{r}_1$ can be selected from a candidate set of size at most $O\big(\log_{1+\varepsilon} k\big)$. For the remaining radii $r_2^*, \ldots, r_k^*$, they lie within the interval $\left[\frac{\varepsilon}{k} r_1^*, r_1^*\right]$. We apply a similar geometric discretization over this interval, producing candidate sets of size $O\left(\log_{1+\varepsilon} k/{\varepsilon}\right)$. Enumerating all such possible radius profiles thus takes time $O\left(\log_{1+\varepsilon}^k k/{\varepsilon}\right).$
\qed
 \end{pf} 
\section{$(2+\varepsilon)$-approximation with time exponential in $k$ and $m$} \label{section2}
In this section, we present a $(2+\varepsilon)$-approximation algorithm with running time $O\left(n^2\log^k(k/\varepsilon)(k + m)^{k + m} \right)$. By Lemma ~\ref{guess}, we can compute a near-optimal radius profile in time $O\left(\log_{1+\varepsilon}^k (k/{\varepsilon})\right)$.
We reduce the problem to a $(k + m)$-sum of radii instance and repeatedly select random points to cover the optimal clusters with enlarged balls.
For the colorful sum of radii instance $I = (P, d, k, \boldsymbol{m})$, the optimal solution consists of $k$ clusters with radii $r_1^*, \ldots, r_k^*$ and $m$ outliers, which can be viewed as balls of radius zero. When a point is randomly selected and the guessed radius matches that of its corresponding optimal cluster, a ball with twice the guessed radius suffices to cover the optimal cluster. If the point is an outlier, its corresponding optimal radius is zero. Repeating this process eventually covers all optimal clusters using a bounded number of enlarged balls. We present the algorithm in Algorithm ~\ref{alg1_beam}.
\begin{algorithm}[htbp] 
\footnotesize
\caption{Iterative Covering}
\label{alg1_beam}
\begin{algorithmic}[1]
\State \textbf{Input:} $P = \{P_i\}_{i=1}^\omega$, $d$, $k$, $\boldsymbol{m}= (m_i)_{i=1}^\omega$, $m$, set of radius profile $\tilde{\mathcal{R}}$
\State Initialize sets $\tilde{\mathcal{R}}', \mathcal{B} \gets \emptyset$, $U \gets (2+\varepsilon) \max_{x,y\in P}d(x,y)$
\For{$(\tilde{r}_1,\dots,\tilde{r}_k) \in \tilde{\mathcal{R}}$}
    \State Pad with $m$ zeros: $(\tilde{r}_1,\dots,\tilde{r}_{k+m}) \gets (\tilde{r}_1,\dots,\tilde{r}_k, 0, \dots, 0)$
    \State $\tilde{\mathcal{R}}' \gets \tilde{\mathcal{R}}' \cup \{(\tilde{r}_1,\dots,\tilde{r}_{k+m})\}$
\EndFor
\For{$(\tilde{r}_1,\dots,\tilde{r}_{k+m}) \in \tilde{\mathcal{R}}'$}
\While{$P \neq \emptyset$}
    \State Pick a point $p \in P$ uniformly at random
    \State Pick an unchosen radius $\tilde{r}_j$ from the set $(\tilde{r}_1,\dots,\tilde{r}_{k+m})$
    \State $\mathcal{B} \gets \mathcal{B} \cup B(p, 2\tilde{r}_j)$, $P \gets P \setminus B(p, 2\tilde{r}_j)$ 
\EndWhile
\If{$|\mathcal{B}| \leq k+m$ \textbf{and }$cost(\mathcal{B}) \leq 2\sum_{i=1}^{k} \tilde{r}_i$} 
\If{$cost(\mathcal{B}) \leq U$}
\State $\mathcal{B}^* \gets \mathcal{B}$, $U \gets cost(\mathcal{B})$
\EndIf
\EndIf
\EndFor
\State \Return $\mathcal{B}^*$
\end{algorithmic}
\end{algorithm}
\begin{theorem}
    There exists a $(2+\varepsilon)$-approximation solution for the colorful sum of radii problem with running time $O(n^2\log^k(k/\varepsilon) (k+m)^{k+m} )$.
\end{theorem}
\begin{pf}
By Lemma \ref{guess}, the size of $\tilde{\mathcal{R}}'$ considered is $O(\log^k(k/\varepsilon))$. For each radius profile in $\tilde{\mathcal{R}}'$, the inner loop iterates $k+m$ rounds to select $k+m$ balls. Suppose that for every selected point $p$ in line 9, we successfully choose $\tilde{r}_j$ in line 10, where $\tilde{r}_j$ is a good approximation of the optimal radius $r_j^*$ of the cluster $C_j^*$ that contains $p$. The probability of correctly selecting the radius is at least $1/(k + m)$, so the probability of correctly all $k + m$ iterations is at least $\geq 1/(k+m)^{k+m}$. To ensure constant success probability, we repeat this procedure $(k+m)^{k+m}$ rounds per profile. Each such procedure performs pick at most $n$ clients, and in each iteration, computing the ball $B(c_j, 2\tilde{r}_j)$ and updating the point set takes $O(n)$ time.
Therefore, the time per profile is $O\left( n^2 (k + m)^{k + m}\right). $ Combining with the size of $\tilde{\mathcal{R}}'$, the total running time is $ O\left( n^2\log^k(k/\varepsilon) \cdot (k + m)^{k + m}\right).$ There must exist at least one successful execution. In this case, for every selected point $p$, the center of the optimal cluster that contains $p$ lies within the ball $B(p, \tilde{r}_j)$. Since $\tilde{r}_j > r_j^*$, the corresponding optimal cluster must be fully contained within the selected ball, i.e., $C_j^* \subseteq B(p, 2\tilde{r}_j)$. The total cost of the solution is
$$
\sum_{i=1}^{k+m} 2\tilde{r}_i = \sum_{i=1}^{k} 2\tilde{r}_i \leq \sum_{i=1}^{k} (1+\varepsilon) \cdot 2r_i^* \leq (2+\varepsilon) \cdot \text{OPT}.
$$
\qed
\end{pf}
\begin{rmk}[Derandomization]\label{remark}
In Algorithm~\ref{alg1_beam}, randomness is introduced in Lines 9 and 10. Specifically, Line 9 randomly selects a point from the set $P$ of at most $n$ points, and Line 10 randomly selects a radius from the set $(\tilde{r}_1, \dots, \tilde{r}_{k+m})$. The total number of possible radii is bounded by $O(\log^k(k/\varepsilon)) \cdot (k+m)$. Therefore, the entire random process can be derandomized by exhaustively enumerating all possible radii.
\end{rmk}
\section{$(7+\varepsilon)$-approximation with time exponential in $k$}
In Section \ref{section2}, we present a $(2+\varepsilon)$-approximation algorithm with running time exponential in both $k$ and $m$. In this section, we propose a $(7+\varepsilon)$-approximation algorithm for the colorful sum of radii problem that removes the exponential dependency on $m$, achieving running time exponential only in $k$. Building upon a constant-factor approximation algorithm for the colorful $k$-center problem, we iteratively select centers and radii to approximate the coverage of the optimal clusters. We formally state the main result of this section as follows:
\begin{theorem}
\label{the1}
 Suppose there exists a $\beta$-approximation algorithm for colorful $k$-center, with running time $T(n,\omega)$, where $\beta$ is a constant and $\omega$ is the number of colors. Then, for any $\varepsilon > 0$, there exists a $(2\beta + 1 + \varepsilon)$-approximation algorithm for the colorful sum of radii problem with running time $O(T(n,\omega)\cdot ((k +m) \cdot \log(k/\varepsilon))^k)$. 
\end{theorem}

Combining Theorem \ref{the1} with the 3-approximation algorithm for colorful $k$-center  ~\cite{jia2022fair} running in $O(n^{\omega^2})$ time, we obtain the main theorem.
\begin{theorem}
    There exists a $(7 + \varepsilon)$-approximation algorithm for the colorful sum of radii problem with running time $O(n^{\omega^2}((k +m)\log(k/\varepsilon))^k)$. 
\end{theorem}

The remainder of this section is dedicated to the proof of Theorem~\ref{the1}.  Section~\ref{sub1} establishes a formal connection between the colorful \( k \)-center problem and the colorful sum of radii.  
Section~\ref{sub2} demonstrates how this relationship can be leveraged to design our main algorithm. Finally, Section~\ref{sub3} provides a theoretical analysis of the algorithm's performance.

\subsection{Bridging colorful $k$-center and colorful sum of radii} \label{sub1}
The colorful \(k\)-center and colorful sum of radii share the same input and solution space; they differ only in their objective functions. While the colorful \(k\)-center aims to minimize the maximum cluster radius, the colorful sum of radii seeks to minimize the total sum of all cluster radii. This structural similarity allows the colorful \(k\)-center algorithm to serve as a useful subroutine in the design of approximation algorithms for the colorful sum of radii.
\begin{lemma}
\label{lemma1}
Let \( I = (P, d, k, \boldsymbol{m}) \) be any instance of the colorful sum of radii, and let \( r_1^* \) denote the largest radius in its optimal solution. Let \( r^* \) denote the optimal radius in the colorful \( k \)-center. Then, \( r^* \leq r_1^* \).
\end{lemma}
\begin{pf}
Let \( \mathcal{O}^*_{\mathrm{SoR}} \) be an optimal solution to the colorful sum of radii with largest radius \( r_1^* \). This solution covers all required points and is therefore a feasible solution to the colorful \(k\)-center with maximum radius \( r_1^* \). Since \( r^* \) is the minimum such radius, we conclude \( r^* \leq r_1^* \).
\qed
\end{pf}

Lemma \ref{lemma1} implies that the optimal radius of the colorful \(k\)-center provides a lower bound on the largest radius in the optimal colorful sum of radii solution. A colorful \(k\)-center algorithm outputs \(k\) centers along with at most \(m\) outliers. Using this solution, we identify the center of the largest optimal cluster and select an appropriate center and radius to cover the corresponding cluster. However, this solution only helps to cover the largest cluster. To generalize this idea, we define a function that tracks the number of points remaining to be covered in each color class. We observe that if the optimal largest cluster is covered, then by removing the covered points, the upper bound of the colorful $k$-center solution is at most twice the next largest radius. We then iteratively invoke the colorful $k$-center algorithm to identify and cover the next optimal cluster, repeating this process until all optimal clusters are covered or the coverage requirements are met.

\begin{figure}[htbp]
    \label{figure1}
    \centering
    \includegraphics[width=0.5\linewidth]{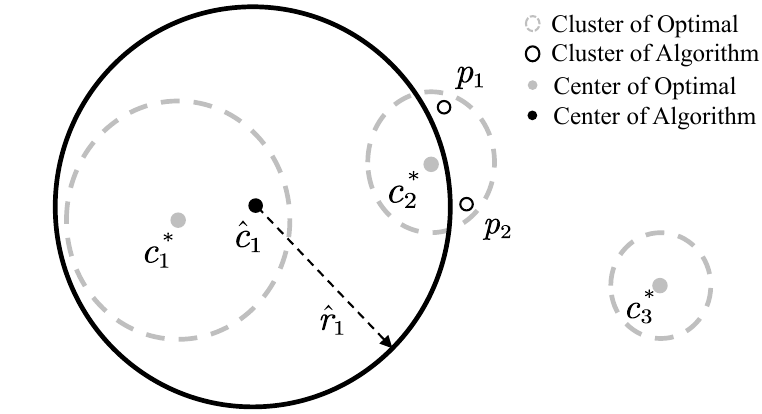} 
    \caption{Given that the first $i{-}1$ optimal clusters are covered, the inclusion of $c_i^*$ within an existing ball ensures that all remaining points are coverable within radius $2r_i^*$}
    \label{fig:your_label}
\end{figure}

Assume that at the start of iteration $i$, the set $ B(\hat{c}_1, \hat{r}_1), \ldots, B(\hat{c}_{i-1}, \hat{r}_{i-1}) $ of selected balls is given. The \textsc{Counting} algorithm then calculates the number of points still required to be covered for each class after removing all points covered by $\bigcup_{h=1}^{i-1} B(\hat{c}_h, \hat{r}_h)$.
\begin{algorithm}[htbp]
\caption{\textsc{Counting}}
\label{alg1}
\textbf{Input:} $P = \{P_i\}_{i=1}^\omega$,   $d$, $\{B(c_h,r_h)\}_{h=1}^{l}$,  $\boldsymbol{\rho}= (\rho_i)_{i=1}^\omega$ \\
\textbf{Output:} $\rho_1',\rho_2',...,\rho_\omega' $
\begin{algorithmic}[1]
\State $\rho_1',\rho_2',...,\rho_\omega' \gets 0$ 
\For{$p \in \bigcup_{h=1}^{l} B(c_h,r_h)$}
\If{$p \in P_i$}
    \State $\rho_i' \gets \hat{\rho}_i +1$
\EndIf
\EndFor
\For{$i = 1 \xrightarrow{} \omega$}
\State $\rho_i' \gets \max\{0,\rho_i\}$
\EndFor
\State \Return $\rho_1',\rho_2',...,\rho_\omega' $
\end{algorithmic}
\end{algorithm}
\begin{lemma}
\label{radius}
Assume that $C_1^*, \ldots, C_{i-1}^*$ are covered by $\bigcup_{h=1}^{i-1} B(\hat{c}_h, \hat{r}_h)$. Let $\boldsymbol{\rho}' \gets \textsc{Counting}(P, d, \{B(\hat{c}_h, \hat{r}_h)\}_{h=1}^{i-1}, \boldsymbol{\rho})$ denote the updated per-class coverage requirements. Update the multi-outlier vector as $\boldsymbol{m}' \gets \boldsymbol{n} - \boldsymbol{\rho}'$ and the point set as $P' \gets P \setminus \bigcup_{h=1}^{i-1} B(\hat{c}_h, \hat{r}_h)$. Then, for the instance $I' = (P', d, k - i, \boldsymbol{m}')$, the optimal largest radius is at most $2 r_i^*$, where $r_i^*$ is the $i$-th largest radius in the optimal solution to $I = (P, d, k, \boldsymbol{m})$.
\end{lemma}
\begin{pf}
Let $\{c^*_h\}_{h=i}^k$ be the remaining optimal centers for clusters $\{C^*_h\}_{h=i}^k$. If $c^*_i$ is not covered by $\bigcup_{h=1}^{i-1} B(\hat{c}_h, \hat{r}_h)$, then all points in $C^*_i \cap P'$ remain assignable to $c^*_i$ within radius $r^*_i$. Since the optimal radii are sorted in non-increasing order, for any uncovered cluster $C^*_h$ with $h > i$, its center $c^*_h$ is also not covered and all points in $C^*_h \cap P'$ can be assigned within radius at most $r^*_i$. If $c^*_i$ is covered, some points in $C^*_i$ may have been removed, but the remaining points can be covered by a ball of radius at most $2r^*_i$ using the triangle inequality. The same argument applies to any covered cluster $C^*_h$ with $h > i$. Figure ~\ref{figure1} illustrates this second case; for clarity, outliers are omitted. Since $c_2^*$ lies inside $B(\hat{c}_1, \hat{r}_1)$, we can construct a feasible solution that covers points $p_1$, $p_2$, and the entire cluster $C_3^*$. The largest radius needed is at most $2r_2^*$. Thus, the optimal largest radius for the modified instance is at most $2r^*_i$.
\qed
\end{pf}

The following corollary follows directly from Lemmas~\ref{lemma1} and~\ref{radius}.
\begin{corollary}
\label{corollary:residual-radius}
Under the assumptions of Lemma~\ref{radius}, let \( \bar{r}_i^* \) denote the optimal radius of the residual colorful \( (k-i) \)-center instance \( \mathcal{I}' = (P', d, k-i, \boldsymbol{m}') \). Then, $\bar{r}_i^* \leq 2r^*_i.$
\end{corollary}
\begin{pf}
By Lemma~\ref{radius}, the instance \( \mathcal{I}' \) admits a feasible solution for the colorful sum of radii problem with maximum radius at most \( 2r^*_i \). Applying Lemma~\ref{lemma1}, we conclude that the optimal radius of the colorful \( (k-i) \)-center problem on \( \mathcal{I}' \) is also at most \( 2r^*_i \), i.e., \( \bar{r}_i^* \leq 2r^*_i \).
\qed
\end{pf}

The colorful $k$-center provides a useful upper bound of $2r_i^*$ for covering the $i$-th largest cluster in the colorful sum of radii solution.
\subsection{Main algorithm} \label{sub2}
In this subsection, we iteratively apply the colorful $k$-center algorithm to construct a feasible solution to the colorful sum of radii problem. The algorithm incrementally builds a set of balls that collectively cover the optimal clusters. Let $\mathcal{B} = \{B(\hat{c}_1, \hat{r}_1), \dots, B(\hat{c}_{i-1}, \hat{r}_{i-1})\}$ be the set of balls selected prior to iteration $i \in [k]$, and let $\mathcal{C} = \{\hat{c}_1, \ldots, \hat{c}_{i-1}\}$ denote the corresponding set of centers, where each center $\hat{c}_j \in \mathcal{C}$ is associated with a radius $\hat{r}_j$. We construct the residual instance $I_i = (P', d, k - i, \boldsymbol{m}')$ by removing the points covered by $\mathcal{B}$ and updating the multi-outlier vector accordingly. We then invoke the colorful $k$-center algorithm (denoted as \textsc{Col-Cen} in Algorithm \ref{alg:colorful-sum-of-radii_beam}) on instance $I_i$, which returns: a set of new centers $\bar{\mathcal{C}}_i$, a uniform covering radius $\bar{r}_i$ for all centers in $\bar{\mathcal{C}}_i$, and a set of uncovered points $\bar{\mathcal{M}}_i$, which are treated as outliers. Each center $c \in \bar{\mathcal{C}}_i$ is associated with radius $\bar{r}_i$, and each outlier $p \in \bar{\mathcal{M}}_i$ is assigned a radius of zero.

Based on above, we define an assignment function $\sigma_i : P \to \mathcal{C} \cup \bar{\mathcal{C}}_i \cup \bar{\mathcal{M}}_i$ as follows. For each point $p \in P$, we assign it to the nearest center $c \in \mathcal{C} \cup \bar{\mathcal{C}}_i \cup \bar{\mathcal{M}}_i$ satisfying $d(p, c) \leq r(c)$, where the radius function $r(\cdot)$ is defined by:
$$
r(c) =
\begin{cases}
\hat{r}_j & \text{if } c = \hat{c}_j \in \mathcal{C}, \\
\bar{r}_i & \text{if } c \in \bar{\mathcal{C}}_i, \\
0 & \text{if } c \in \bar{\mathcal{M}}_i.
\end{cases}
$$
Then, we assign $p$ to the nearest center $c \in \mathcal{C} \cup \bar{\mathcal{C}}_i \cup \bar{\mathcal{M}}_i$ that satisfies $d(p, c) \leq r(c)$, i.e.,
$$
\sigma_i(p) = \arg\min_{\substack{c \in \mathcal{C} \cup \bar{\mathcal{C}}_i \cup \bar{\mathcal{M}}_i \\ d(p, c) \leq r(c)}} d(p, c),
$$ Based on this assignment function, we determine which center covers the optimal cluster center $c_i^*$, and then enlarge the radius accordingly to ensure that the entire cluster $C_i^*$ is covered.
\begin{itemize}
    \item If \( \sigma_i(c_i^*) \in \mathcal{C} \), we extend the radius of the corresponding existing center in \( \mathcal{C} \) by an additive factor of \( (2\beta + 1)\tilde{r}_i \). Additionally, we randomly select a point \( x \in P \setminus \bigcup_{B \in \mathcal{B}} B \) to serve as a dummy center \( \hat{c}_i = x \) with radius \( \hat{r}_i = 0 \).
    \item If \( \sigma_i(c_i^*) \in \bar{\mathcal{C}}_i \), we add \( \hat{c}_i = \sigma_i(c_i^*) \) to \( \mathcal{C} \) with the radius \( \hat{r}_i = (2\beta + 1) \tilde{r}_i \).
    \item If \( \sigma_i(c_i^*) \in \bar{\mathcal{M}}_i \), we add \( \hat{c}_i = \sigma_i(c_i^*) \) to \( \mathcal{C} \) with radius \( \hat{r}_i = (2\beta + 1) \tilde{r}_i \).
\end{itemize}
\begin{algorithm}[htbp]
\caption{Main algorithm}
\label{alg:colorful-sum-of-radii_beam}
\begin{algorithmic}[1]
\footnotesize
\State \textbf{Input:} $P = \{P_i\}_{i=1}^\omega$, $d$, $k$, $\boldsymbol{m}= (m_i)_{i=1}^\omega$, $\boldsymbol{\rho}= (\rho_i)_{i=1}^\omega$, set of radius profile $\tilde{\mathcal{R}}$,
\State \textbf{Initialize:} $\mathcal{B},\mathcal{C},\boldsymbol{n}' \gets \emptyset$, $U \gets (7+\varepsilon) \max_{x,y\in P} d(x,y)$
\For{$(\tilde{r}_1,...,\tilde{r}_k) \in \tilde{\mathcal{R}}$}
\For{$i = 1$ to $k$}
\State $P_i \gets P\setminus\bigcup_{B\in \mathcal{B}}B$
\State Compute class-wise counts vector $\boldsymbol{n}'$ from $P_i$
\State $\boldsymbol{m}' \gets \boldsymbol{n}' - \textsc{Counting}(P, d, \mathcal{B}, \boldsymbol{\rho})$
\If{$\boldsymbol{m}' = \boldsymbol{n}'$}
\State \textbf{break}
\EndIf
\State $(\bar{\mathcal{C}}_i,\bar{\mathcal{M}}_i, \bar{r}_i) \gets \textsc{Col-Cen}(P_i, d, k-i, \boldsymbol{m}')$
\State Construct assignment $\sigma_i$ from $P$ to $\mathcal{C} \cup \bar{\mathcal{C}}_i \cup \bar{\mathcal{M}}_i$
\If{$\sigma_i(c^*_i) = \hat{c}_j \in \mathcal{C}$} 
\State $\hat{r}_j \gets \hat{r}_j + (2\beta + 1)\tilde{r}_i$, $\mathcal{B}[j] \gets B(\hat{c}_j,\hat{r}_j)$
\State Pick a random point $x \in P \setminus \bigcup_{B \in \mathcal{B}} B$ and let $\hat{c}_i \gets x$, $\hat{r}_i \gets 0$
\State $\mathcal{B} \gets \mathcal{B} \cup   B(\hat{c}_i,\hat{r}_i)$
\ElsIf{$\sigma_i(c^*_i) = \bar{c} \in \bar{\mathcal{C}}_i$}
\State $\hat{c}_i \gets \bar{c}$, $\hat{r}_i \gets (2\beta + 1)\tilde{r}_i$.
\State  $\mathcal{C} \gets \mathcal{C} \cup \{\hat{c}_i\}$, $\mathcal{B} \gets \mathcal{B} \cup   B(\hat{c}_i,\hat{r}_i)$
\ElsIf{$\sigma_i(c^*_i) = \bar{o}\in \bar{\mathcal{M}}_i$}
\State $\hat{c}_i \gets \bar{o}$, $\hat{r}_i \gets  (2\beta + 1)\tilde{r}_i$.
\State $\mathcal{C} \gets \mathcal{C} \cup \{\hat{c}_i\}$, $\mathcal{B} \gets \mathcal{B} \cup   B(\hat{c}_i,\hat{r}_i)$
\EndIf
\EndFor
\If{$\mathcal{B}$ is feasible and $cost(\mathcal{B}) \leq U$}
\State $\mathcal{B}^* \gets \mathcal{B}$, $U \gets cost(\mathcal{B})$
\EndIf 
\EndFor
\State \Return $\mathcal{B}^*$
\end{algorithmic}
\end{algorithm}

In the main algorithm, for each radius profile, we perform at most $k$ iterations to find a feasible solution. At iteration $i$, before selecting the next optimal cluster $C_i^*$, we update and examine the multi-outlier vector $\boldsymbol{m}'$. Let $\boldsymbol{n}'$ denote the counts of remaining points per class in the residual set $P_i$. If $\boldsymbol{m}' = \boldsymbol{n}'$, all remaining points can be treated as outliers, indicating that a feasible solution has been found and the inner loop terminates. Otherwise, the algorithm proceeds to cover the next cluster. After iterating over all radius profiles, the best solution is returned.
\subsection{Analysis} \label{sub3}
We analyze the algorithm under the favorable event $\mathcal{E}$, defined as: at iteration $i$, under the assignment function $\sigma_i$, the $i$-th optimal cluster center $c_i^*$ is correctly assigned to its corresponding center in the current solution. We will later prove that $\mathcal{E}$ occurs with constant probability.
\begin{lemma}
(Conditioned on event \( \mathcal{E} \)) 
Assume that $\tilde{r}_1, \ldots, \tilde{r}_k$ is near-optimal, and $C_1^*,...,C_i^*$ are  covered by $\{ B(\hat{c}_h, \hat{r}_h) \}_{h=1}^{i-1}$. In the instance $I_i = \big(P_i, d, k - i, \boldsymbol{m}'\big)$, the subroutine \textsc{Col-cen} returns a radius \( \bar{r}_i \) such that \( \bar{r}_i \leq 2\beta \tilde{r}_i \).
\end{lemma}

\begin{pf}
By Corollary \ref{corollary:residual-radius}, the optimal radius \( \bar{r}_i^* \) for colorful $(k-i)$-center in instance \( I' \) satisfies \( \bar{r}_i^* \leq 2r_i^* \). Since the algorithm returns a \( \beta \)-approximate radius, we have $\bar{r}_i \leq \beta \bar{r}_i^* \leq \beta r_i^* \leq 2\beta \tilde{r}_i.$
\qed
\end{pf}
Hence, we can upper bound the cluster radius by $\bar{r}_i \leq 2\beta \tilde{r}_i$. We then show that, conditioned on the event $\mathcal{E}$, the first $i-1$ optimal clusters are fully covered by the previously selected balls $\{ B(\hat{c}_h, \hat{r}_h) \}_{h=1}^{i-1}$ at iteration $i$.
\begin{lemma}\label{lem:covering}
(Conditioned on event \( \mathcal{E} \)) 
Let \( \tilde{r}_1, \ldots, \tilde{r}_k \) be a near-optimal radius profile, and let $\mathcal{B} = \{ B(\hat{c}_1, \hat{r}_1), \ldots, B(\hat{c}_k, \hat{r}_k) \}$
denote the set of balls returned by the algorithm. Then:
\begin{enumerate}
    \item For each \( j \in [k] \), there exists \( \ell \in [k] \) such that $ C_j^* \subseteq B(\hat{c}_\ell, \hat{r}_\ell).$
    \item For every \( \ell \in [k] \) with \( \hat{r}_\ell > 0 \), there exists \( j \in [k] \) such that $C_j^* \subseteq B(\hat{c}_\ell, \hat{r}_\ell).$
\end{enumerate}
\end{lemma}
\begin{pf}
We proceed by induction on the iteration number \( i = 0, \ldots, k \). Let \( \mathcal{B}_i \) denote the set of balls after iteration \( i \). We maintain the following inductive invariants:
\begin{itemize}
    \item[(I1)] For each \( j \in [i] \), there exists \( \ell \in [i] \) such that \( C_j^* \subseteq B(\hat{c}_\ell, \hat{r}_\ell) \).
    \item[(I2)] For each \( \ell \leq i \) with \( \hat{r}_\ell > 0 \), there exists \( j \in [i] \) such that \( C_j^* \subseteq B(\hat{c}_\ell, \hat{r}_\ell) \).
\end{itemize}

At iteration $i = 0$, since no balls have been selected and no clusters have been covered, the statement trivially holds. Assume the invariants hold at step \( i-1 \). At step \( i \), consider the \( i \)-th optimal cluster \( C_i^* \) centered at \( c_i^* \), and let \( \sigma_i(c_i^*) \) denote the guessed assignment. We analyze three cases:

\textbf{Case 1:} \( \sigma_i(c_i^*) = \hat{c}_j \) for some \( j \leq i - 1 \).  
If \( \hat{r}_j = 0 \), then for each \( p \in C_i^* \),
\[
d(p, \hat{c}_j) = d(p, c_i^*) \leq r_i^* \leq 2(\beta + 1) \tilde{r}_i \leq  2(\beta + 1) \tilde{r}_i +\hat{r}_j,
\]
implying \( C_i^* \subseteq B(\hat{c}_j, \hat{r}_j + 2(\beta + 1) \tilde{r}_i) \).  
If \( \hat{r}_j > 0 \), then
\[
d(p, \hat{c}_j) \leq d(p, c_i^*) + d(c_i^*, \hat{c}_j) \leq r_i^* + \hat{r}_j \leq \tilde{r}_i + \hat{r}_j \leq  2(\beta + 1) \tilde{r}_i +\hat{r}_j .
\]
In both cases, the cluster \( C_i^* \) is covered by the ball $B(\hat{c}_j, \hat{r}_j + 2(\beta + 1) \tilde{r}_i)$.

\textbf{Case 2:} \( \sigma_i(c_i^*) = \bar{c} \in \bar{\mathcal{C}_i} \).  
We select \( \hat{c}_i = \bar{c} \) with radius \( \hat{r}_i = (2\beta + 1)\tilde{r}_i \). Then for each \( p \in C_i^* \),
\[
d(p, \hat{c}_i) \leq d(p, c_i^*) + d(c_i^*, \bar{c}) \leq r_i^* + \bar{r}_i \leq (2\beta + 1)\tilde{r}_i.
\]
The final inequality follows from the bound $\bar{r}_i \leq 2\beta\tilde {r}_i$.

\textbf{Case 3:} \( \sigma_i(c_i^*) = \bar{o} \in \bar{\mathcal{M}}_i \).  
We choose $\hat{c}_i = \bar{o}$ and set $\hat{r}_i = (2\beta + 1)\tilde{r}_i$. Since the outlier set is assigned to itself, we have $d(c_i^*, \bar{o}) = 0$, and thus for any point $p \in C_i^* $ 
\[
d(p, \hat{c}_i) = d(p, c_i^*) + d(c_i^*, \bar{o}) \leq r_i^* \leq \tilde{r}_i \leq (2\beta + 1)\tilde{r}_i.
\]

In all cases, invariants (I1) and (I2) are maintained.
\qed
\end{pf}
Under the assumption that event $\mathcal{E}$ occurs, Lemma \ref{lem:covering} guarantees that Algorithm \ref{alg:colorful-sum-of-radii_beam} returns a feasible solution. We analyze its approximation ratio as follows.
\begin{lemma}\label{ratio}
(Conditioned on event \( \mathcal{E} \)) Let \( \tilde{r}_1, \ldots, \tilde{r}_k \) be a near-optimal radius profile, and
$\mathcal{B} = \{ B(\hat{c}_1, \hat{r}_1), \ldots, B(\hat{c}_k, \hat{r}_k) \}$ denote the set of balls returned by the algorithm. The algorithm returns a feasible solution whose total cost is at most \( (2\beta + 1 + \varepsilon) \cdot \mathrm{OPT} \).
\end{lemma}

\begin{pf}
We prove by induction that for all \( i \in [1,k] \),
\[
\sum_{j=1}^i \hat{r}_j \leq (2\beta + 1) \sum_{j=1}^i \tilde{r}_j.
\]
The base case \( i = 0 \) is trivial. Assume the bound holds for \( i-1 \). At iteration \( i \), by the algorithm’s selection rule, we have \( \hat{r}_i \leq (2\beta + 1)\tilde{r}_i \), yielding the claim. Since \( \sum_{j=1}^{k} \tilde{r}_j \leq (1+\varepsilon)\mathrm{OPT} \), the total cost satisfies:
\[
\sum_{j=1}^k \hat{r}_j \leq (2\beta + 1)\sum_{j=1}^k \tilde{r}_j \leq (2\beta + 1 +\varepsilon) \mathrm{OPT}.
\]
\qed
\end{pf}

In the inner loop, at each iteration \( i \), the probability that the algorithm correctly guesses the assignment of \( c^*_i \) is at least $1/k+m$. Since there are at most \( k \) iterations, the event \( \mathcal{E} \) occurs with probability at least \( 1/({k+m})^k \). Repeating the algorithm independently \( (k+m)^k \) times yields, with constant probability, a feasible solution satisfying the approximation guarantees. Since each invocation of the colorful $k$-center algorithm takes \( T(n, \omega) \) time, the overall running time is $O(T(n,\omega)\cdot ((k +m) \cdot \log(k/\varepsilon))^k)$. This process can be derandomized in a manner similar to that described in Remark \ref{remark}.

\end{document}